\documentclass[aps,prd,eqsecnum,11pt,showpacs,superscriptaddress,nofootinbib]{revtex4-2}
\usepackage{geometry}                		% See geometry.pdf to learn the layout options. There are lots.
\usepackage{graphicx}				% Use pdf, png, jpg, or epsï¿½ with pdflatex; use eps in DVI mode
\usepackage{amssymb}
\usepackage{graphicx}% Include figure files
\usepackage{amssymb,amsmath}
\usepackage{amsfonts,mathrsfs}
\newcommand{\la}{\langle}
\newcommand{\ra}{\rangle}
\newcommand{\be}{\begin{equation}}
\newcommand{\ee}{\end{equation}}
\newcommand{\bea}{\begin{eqnarray}}
\newcommand{\eea}{\end{eqnarray}}
\newcommand{\bes}{\begin{subequations}}
\newcommand{\ees}{\end{subequations}}
\newcommand{\w}{\omega}

\begin{document}

\title{ The fine structure of the peaks of the correlation function\\  in acoustic black holes: a complete analytical model}

\author{Paul~R.~Anderson}
\email{anderson@wfu.edu}
\affiliation{Department of Physics, Wake Forest University, Winston-Salem, North Carolina 27109, USA}
\author{Roberto~Balbinot}
\email{Roberto.Balbinot@bo.infn.it}
\affiliation{Dipartimento di Fisica dell'Universit\`a di Bologna and INFN sezione di Bologna, Via Irnerio 46, 40126 Bologna, Italy}
\author{Richard~A.~Dudley}
\email{Richard.A.Dudley@charlotte.edu}
\affiliation{Department of Physics and Optical Sciences, University of North Carolina at Charlotte, Charlotte, North Carolina 28223, USA}
\author{Alessandro~Fabbri}
\email{afabbri@ific.uv.es}
\affiliation{Departamento de F\'isica Te\'orica and IFIC, Universidad de Valencia-CSIC, Calle Dr. Moliner 50, 46100 Burjassot, Spain}
\author{Amanda~Peake}
\email{peakar21@wfu.edu}
\affiliation{Department of Physics, Wake Forest University, Winston-Salem, North Carolina 27109, USA}
%\affiliation{Centro Studi e Ricerche E. Fermi, Piazza del Viminale 1, 00184 Roma, Italy}
%\affiliation{Dipartimento di Fisica dell'Universit\`a di Bologna and INFN sezione di Bologna, Via Irnerio 46, 40126 Bologna, Italy}
%\affiliation{Laboratoire de Physique Th\'eorique, CNRS UMR 8627, B\^at. 210, Universit\'e Paris-Sud 11, Univ. Paris-Saclay, 91405 Orsay Cedex, France}
\author{Daniel~Pe\~{n}alver}
\email{Daniel.Penalver@ific.uv.es}
\affiliation{Departamento de F\'isica Te\'orica and IFIC, Universidad de Valencia-CSIC, Calle Dr. Moliner 50, 46100 Burjassot, Spain}

%\affiliation{INFN sezione di Bologna, Via Irnerio 46, 40126 Bologna, Italy}
%\affiliation{Centro Studi e Ricerche E. Fermi, Piazza del Viminale 1, 00184 Roma, Italy}

\begin{abstract}

The detailed structure of the peaks appearing in the density-density correlation function for an acoustic black hole formed by a Bose-Einstein condensate is
analytically discussed for a particular, but physically meaningful, sound velocity profile that allows the field modes to be exactly computed.

%  A complete set of exact analytic solutions to the mode equation are found in the region exterior to the acoustic horizon
%for 1D Bose-Einstein condensate (BEC) acoustic black holes.
%%with a particular type of sound speed profile.
% From these, analytic expressions for the scattering coefficients and graybody factor are obtained. %The results are used
% to verify previous predictions regarding the behaviors of the scattering coefficients and graybody factor in the low frequency limit.
 \end{abstract}

\maketitle

\section{Introduction}

Since the original proposal of 2008 \cite{paper1}, analysis of correlation functions for acoustic black holes (BHs) has become the standard tool in theoretical and experimental investigations of the Hawking effect in these systems.
The analysis of \cite{paper1} predicted a peak in the in-out (one point inside the horizon, the other outside) density-density correlation function of a Bose-Einstein condensate (BEC) sonic BH located along the geometrical optics trajectories of the Hawking particle and its partner.

This study, performed by the methods of Quantum Field Theory (QFT) in curved spacetime (see for example \cite{bd, fu, pt}), neglected the backscattering of the modes caused by the curvature of the acoustic spacetime, so the resulting theory was simply that of a massless conformally invariant
scalar field in two dimensions which can be solved exactly. When backscattering is included one has to resort to numerical investigations which showed the existence of two more peaks in the correlation function of minor intensity compared to the previous one \cite{paper2, rpc, mp, paper2011}.

Steinhauer and coworkers in a series of experiments performed with BECs \cite{jeff2016, jeff2019, jeff2021} confirmed the existence of the main peak located, within the sensitivity of their measurements, along the predicted trajectories. At present, this is the best experimental evidence of Hawking radiation in acoustic BHs. Because of the weakness of the corresponding signals, the two minor peaks have so far not been detected.

Without underestimating the importance of numerical work, we feel that analytically solvable models can significantly increase our understanding of the fine structure of the peaks in the correlation function and can serve as a guide for future experiments.

In this paper we propose such a model characterised by a peculiar, but physically meaningful, sound velocity profile which allows the modes of the field to be exactly calculated even in the presence of backscattering and definite analytical predictions about the peaks and their features to be made. 

\section{The sound velocity profile}

Under the hydrodynamic approximation,
%, where deviations in the speed of sound and density occur on scales larger than the healing length, $\xi=\frac{\hbar}{mc}$, of the condensate, 
the quantum phase fluctuation $\hat{\phi}$ of a BEC satisfies an equation which is equivalent
% that quantify the quantum fluctuations on top of the classical state can be described by 
%\be
%\left[-\left(\partial_t+\vec{\nabla}\vec{v_0} \right)\frac{n}{mc^2}\left(\partial_t+\vec{v_0}\vec{\nabla}\right)+\vec{\nabla}\frac{n}{m}\vec{\nabla}\right]{\hat\phi}=0\label{KGE}
%\ee
%where $n$ is the density of the condensate (constant, for the model considered) , and $m$ is the mass of an atom. Eq. \eqref{KGE} is equivalent 
to the Klein-Gordon equation for a massless and minimally coupled scalar field \cite{Barcelo:2005fc}  \be\label{kg} 
\Box \hat\phi = \frac{1}{\sqrt{\text{-} g}}\ \partial_\mu \left(\sqrt{\text{-}g}\ g^{\mu \nu} \partial_\nu \hat\phi \right)=0\ 
\ee
in a fictitious curved spacetime described by the metric
\be \label{am} ds^2=\frac{n}{mc}[-c^2dT^2+(d\vec x - \vec v dT)(d\vec x - \vec v dT)]\ , \ee
where $\vec v$ and $n$ are the speed and the density of the condensate, $c$ is the speed of sound and $m$ the mass of an atom.

We consider a  1D BEC \cite{ps, ms} moving at a negative constant velocity $-v_0\ (v_0>0)$ along the $\-\hat{x}$ direction. The speed of sound $c=c(x)$
% is modified along the $\hat{x}$-axis via a Feshbach resonance (see \cite{ps}) via a background magnetic fields so that 
in the limit that $x\to \infty$ reaches an asymptotic value $c(x)\to c_R>v_0$ and as $x\to -\infty$, $c(x)\to c_L<v_0$. By setting $c(0)=v_0$, the system is an acoustic analog of an asymptotically flat black hole with a horizon at $x=0$.  
The acoustic metric (\ref{am}) is then %The acoustic metric for this curved spacetime can be put in the form 
\be\label{mas}
ds^2=\frac{n}{mc}\left[-(c^2-v_0^2)dT^2 + 2v_0 \, dT \, dx +dx^2 + dx_{\perp}^2 \right]
\ , \ee
where $x_{\perp}$ denote the directions perpendicular to $x$.
It is both instructive and useful to use a Schwarzschild time $t$ defined by
%  Outside the horizon
\be
    t=T-\int^x dy \frac{v_0}{c(y)^2-v_0^2} \;, \ee
%and inside the horizon
%\begin{equation}
%  t=T-\int_{x_2}^x dy \frac{v_0}{c(y)^2-v_0^2} +a  \;,
%\end{equation}
%\ees
%where $a$ is an arbitrary constant which is discussed in section \ref{Sec:Numerical Details}.  
resulting in the line element 
\be
ds^2=\frac{n}{m}\left[-\frac{c^2-v_0^2}{c}dt^2+\frac{c}{c^2-v_0^2}dx^2+\frac{dx_\perp^2}{c} \right]\ .
\ee

.

%There are two sound speed profiles that we use in this paper.  The first is 
%the profile used in~\cite{paper2013} (originally taken to match the one used in~\cite{paper2}).  It is 
%\bea  c(x)&=&\sqrt{c_L^2 + \frac{1}{2}(c_R^2-c_L^2)\left[1+\frac{2}{\pi}\tan^{-1}\left(\frac{x+b}{\sigma_v}\right)\right]}  \;, \nonumber \\
%      b &=& \sigma_v \tan \left[\frac{\pi}{c_R^2-c_L^2} \left( v_0^2 - \frac{1}{2} (c_R^2+c_L^2)\right) \right] \;, \label{c-used} \eea
%where $\sigma_v$ is an arbitrary positive constant. \

%\subsection{Analytic mode solutions}

The analytic sound speed profile we will use is a $C^1$ function at the horizon, $x = 0$.   
To motivate it, we rewrite (\ref{kg}) in the metric (\ref{mas})  and let $\phi(t, x)=e^{-i\omega t}\varphi(x)$.  The resulting equation for $\varphi$ is\footnote{For our one dimensional flow, the relevant modes of $\hat \phi$ are those which are constant along $x_\perp$. }
\be
\frac{d}{dx}\left[\left(\frac{c^2-v_0^2}{c^2} \right) \frac{d\varphi }{dx} \right]+ \frac{\omega^2}{c^2-v_0^2}\varphi=0  \;.
\ee
%For the models we consider, $c > v_0$ for $x > 0$ and $c < v_0$ for $x < 0$, while $c = v_0$
%at $x = 0$ which is the acoustic horizon.

It is useful to introduce the spatial coordinate $z$ via 
\be  dx = \left(1-\frac{v_0^2}{c(z)^2} \right) \,dz \ . \label{uu} \ee
%which, looking at the definition of the tortoise coordinate $x^*$ in (\ref{tc}), satisfies $dx^*=\frac{dz}{c(z)}$. 
This transformation works both for $x > 0$ and for $x < 0$, but not simultaneously.  For both cases, $x = 0$ corresponds to $z = -\infty$.  However, for $x >0$, $x = +\infty$ corresponds to $z = +\infty$, while for $x < 0$, $x = -\infty$ corresponds to $z = + \infty$.
The resulting mode equation for both $x > 0$ and $x < 0$ is
\be
\left[\frac{\partial ^2}{\partial z^2} \varphi + \frac{\w^2}{c(z)^2 } \right]\varphi=0  \;.  \label{varphi-z-eq}
\ee

\begin{figure}[h]
\centering 
\includegraphics[width=3.5in] {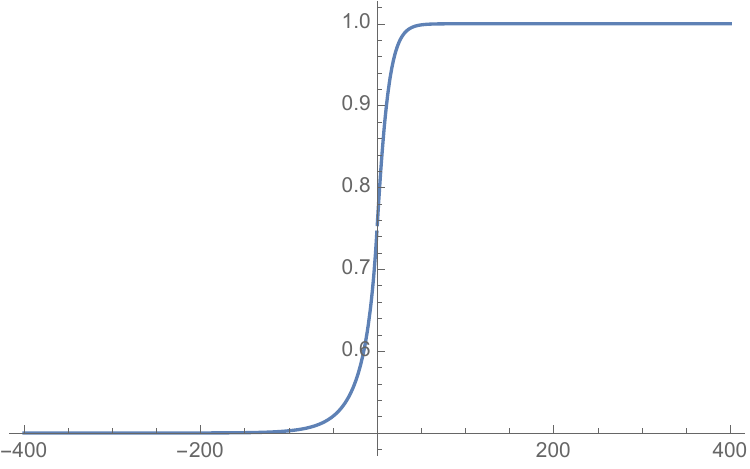}
\caption{Plot of the profile (\ref{c-x-lt-0}),(\ref{c-x-gt-0}) for $c_L=\frac{1}{2}, v_0=\frac{3}{4}, c_R=1$ and $\kappa=0.0185617$.}
\label{figuno}
\end{figure}

Analytic solutions to this equation can be found using the speed of sound profile \cite{fba2016}
%~\cite{fba2016} 
\be
c  =[A+B \tanh (k z)]^{-1/2} \;, \label{c-profile-1} \\
\ee
where ${A}$,  ${B}$ and  $k$ are constants. 
In \cite{fba2016},
%~\cite{fba2016}, 
only the case $x \ge 0$ was considered.  However, this profile can also be used for $x \le 0$, but with different values for the constants $A$ and $B$.
For $x \ge 0$, we choose $A$ and $B$ so that $c \to c_R > v_0$ in the limit $x \to \infty$ and $c = v_0$ at $x = 0$.  Evaluating~\eqref{c-profile-1} in these limits gives
\bea  A_R + B_R &=& \frac{1}{c_R^2}\ ,   \nonumber \\
      A_R - B_R &=& \frac{1}{v_0^2} \;. \label{ARBR} \eea
For $x \le 0$, we choose $A$ and $B$ so that $c \to c_L < v_0$ in the limit $x \to -\infty$ and $c = v_0$ at $x = 0$.  In this case
\bea  A_L + B_L &=& \frac{1}{c_L^2} \ ,  \nonumber \\
      A_L - B_L &=& \frac{1}{v_0^2} \;. \label{ALBL} \eea
Substituting~\eqref{c-profile-1} into~\eqref{uu} and using~\eqref{ARBR} gives for $x \ge 0$
%\bes 
\be x = \frac{1}{2 k} \left(1 - \frac{v_0^2}{c_R^2} \right) \log \left( e^{2 k z} + 1 \right) \;. \label{x-z-right}  \ee
For $x \le 0$, using~\eqref{ALBL} gives
\be x = \frac{1}{2 k} \left(1 - \frac{v_0^2}{c_L^2} \right) \log \left( e^{2 k z} + 1 \right) \;. \label{x-z-left}  \ee %\ees
Then in terms of $x$, the sound speed profile for $x >0$ is
\bes \be c^{ext} =  \frac{c_R}{\sqrt{1 + \left(\frac{c_R^2}{v_0^2} - 1 \right) \exp \left[-\frac{2 \frac{c_R^2}{v_0^2} k x}{(\frac{c_R^2}{v_0^2} - 1)}\right]}}  \;, \label{c-x-gt-0}
\ee
and the one for $x < 0$ is
\be c^{int} =  \frac{c_L}{\sqrt{1 - \left(1 -\frac{c_L^2}{v_0^2} \right) \exp \left[-\frac{2 k \frac{c_L^2}{v_0^2} x}{\left(\frac{c_L^2}{v_0^2}-1\right)}\right]}} \;. \label{c-x-lt-0}
\ee  \ees  
The surface gravity is given by
\be \kappa =  \left. \frac{d c}{dx} \right \vert_{x=0} = k \, v_0\ .  \ee
The plot of the profile is given Fig (\ref{figuno}).

As one can see, the profile mimics typical ones adopted for acoustic BHs with two asymptotic  regions that are almost flat with, one subsonic ($x \to  + \infty$), and the other supersonic ($x \to - \infty$), where $c(x) = c_R$ and $c(x) =c_L$ respectively.
One should remark that since this profile is $C^1$ at the horizon, the effective potential for the modes is continuous there (see Eqs. (2.9), (2.4) of Ref \cite{paper-renaud}).
For this profile the modes can be obtained analytically in terms of hypergeometric functions. This construction has been given in Ref. \cite{fba2016} for the subsonic ($x>0$) region. In Appendix A we summarise it and extend it to the supersonic region ($x<0$). 

\section{ Two-point function and  density-density correlation function}

Following \cite{paper2013}
%~\cite{paper2013} 
one can write the equal lab time density-density correlation function as  
\be
G_2\left(T, x , x^\prime\right)=\lim_{T^
\prime \to T}\left<\hat{n}(T, x)\hat{n}(T^\prime, x^\prime) \right>\ ,
\ee
where the density can be written in terms of the phase operator using the hydrodynamic approximation 
\be
\hat{n}\simeq -\frac{\hbar n}{mc^2}\left[\vec{v_0}\vec{\nabla}\hat{\phi}+\partial_T  \hat{\phi}\right]\ .
\ee
The density-density correlation function for our 1D condensate then becomes
\bes \bea
G_2\left(T, x , x^\prime\right) &=& \lim_{T^\prime\to T} \frac{\hbar^2 n^2}{m^2c^2(x)c^2(x^\prime)}
D\left<  \left\{\hat{\phi}(T, x), \hat{\phi}(T^\prime, x^\prime)\right\}\right>\ , 
\\ 
D &\equiv& \partial_T\partial_{T^\prime}-v_0\partial_x\partial_{T^\prime}-v_0\partial_T\partial_{x^\prime}+v_0^2\partial_x\partial_{x^\prime} \ .
\eea \label{G2-eq} \ees
%Substituting (\ref{fd}), using (\ref{Kruskal-modes}) and (\ref{alphas-betas}), and integrating the integral over $\w_K$ first as was done in~\cite{paper2013}, one finds that

\begin{figure}[h]
\centering 
\includegraphics[width=4.5in] {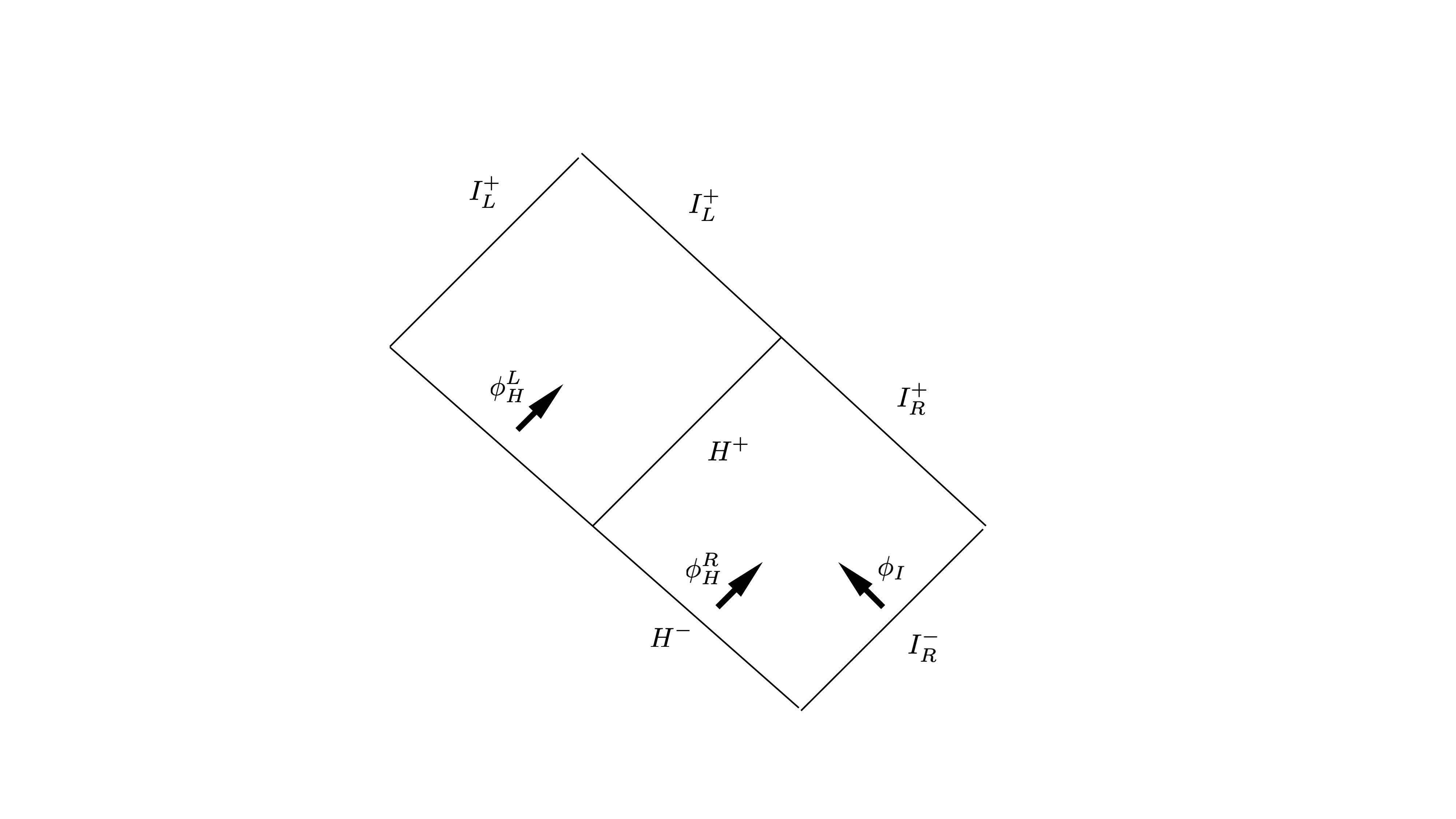}
\caption{Penrose diagram of our acoustic black hole model and origin of the modes $\phi_I, \phi_H^R, \phi_H^L$.}
\label{figdue}
\end{figure}

The two-point function of $\hat\phi$ is computed in the Unruh state $|U\rangle$ \cite{unruh76} , which reproduces Hawking radiation at late times, and takes the form \cite{paper2013}\footnote{Strictly speaking there should be lower limit cutoffs on these integrals because there are infrared divergences in their integrands.  However, the density-density correlation function is independent of these divergences~\cite{rigorous-results}.}

\bes  \bea & &\la  U|\left\{\hat{\phi}(T, x), \hat{\phi}(T^\prime, x^\prime) \right\} |U\ra = I + J \ ,\label{tpf} \\ \label{Ieqa}
  I &=& \int_0^\infty d \omega \frac{1}{\sinh\left(\frac{\pi \omega}{\kappa}\right)} \left\{ \phi^{L}_H(\omega,T,x) \, \phi^{R}_H(\omega,T',x') +
   \phi^{L *}_H(\omega,T,x) \, \phi^{R *}_H(\omega,T',x') \right. \nonumber \\
   & & \;\;\; \left. + \phi^{R}_H(\omega,T,x) \, \phi^{L}_H(\omega,T',x') +
   \phi^{R *}_H(\omega,T,x) \, \phi^{L *}_H(\omega,T',x') \right. \nonumber \\
   & & \;\; \left.  + \cosh \left(\frac{\pi \omega}{\kappa}\right) \left[ \phi^{L}_H(\omega,T,x) \, \phi^{L *}_H(\omega,T',x')
   + \phi^{L *}_H(\omega,T,x) \, \phi^{L}_H(\omega,T',x') \right. \right. \nonumber \\
  & & \;\;\; \left. \left. + \phi^{R}_H(\omega,T,x) \, \phi^{R *}_H(\omega,T',x') + \phi^{R *}_H(\omega,t,x) \, \phi^{R }_H(\omega,T',x') \right]
   \right\} \ ,  \\
  J &=& \int_0^\infty d \omega \, \left[ \phi^{R}_I(\omega,T,x) \, \phi^{{R} \, *}_I(\omega,T',x')
   +  \phi^{{R} \, *}_I(\omega,T,x) \, \phi^{R}_I(\omega,T',x') \right] \;. \label{Jdef} 
   \eea \label{I-J} \ees  
   The modes $\phi_I, \phi_H^R, \phi_H^L$ entering eqs. (\ref{Ieqa}) and (\ref{Jdef}) are  defined so that (see the Penrose diagram in Fig. (\ref{figdue}))
the mode $\phi_I$ is purely ingoing on past null infinity $I^-_R$, where it has the form (in $t,z$ coordinates used in the previous section, see \cite{fba2016} and Appendix A)
\be \label{opi}
\phi_I = \sqrt{\frac{mc_R}{4\pi\omega n_{1D}\hbar}}e^{-i\omega(t+\frac{z}{c_R})}\ ,
\ee
where $n_{1D}=nL_\perp^2$ is the 1D density and the size of the transverse dimension is $L_\perp$.  We assume that  $L_\perp \ll \frac{\hbar}{mc}$, so that excitations with transverse momenta are frozen.  The mode 
$\phi_H^R$ is purely outgoing on the $x>0$ portion of the past horizon $H^-$  where it has the form
\be \label{opr}
\phi_H^R = \sqrt{\frac{mv_0}{4\pi\omega n_{1D}\hbar}}e^{-i\omega(t-\frac{z}{v_0})}\ .
\ee
Finally $\phi_H^L$ is purely outgoing on the $x<0$ portion of $H^-$ where it has the form
\be \label{opl}
\phi_H^L = \sqrt{\frac{mv_0}{4\pi\omega n_{1D}\hbar}}e^{i\omega(t-\frac{z}{v_0})}\ .
\ee

These modes along with the scattering coefficients associated with them are shown schematically in Figs. (\ref{figtre}), (\ref{figquattro}), (\ref{figcinque}).  Exact analytic expressions for them are given in Appendix A.

\begin{figure}[h]
\centering 
\includegraphics[width=4.5in] {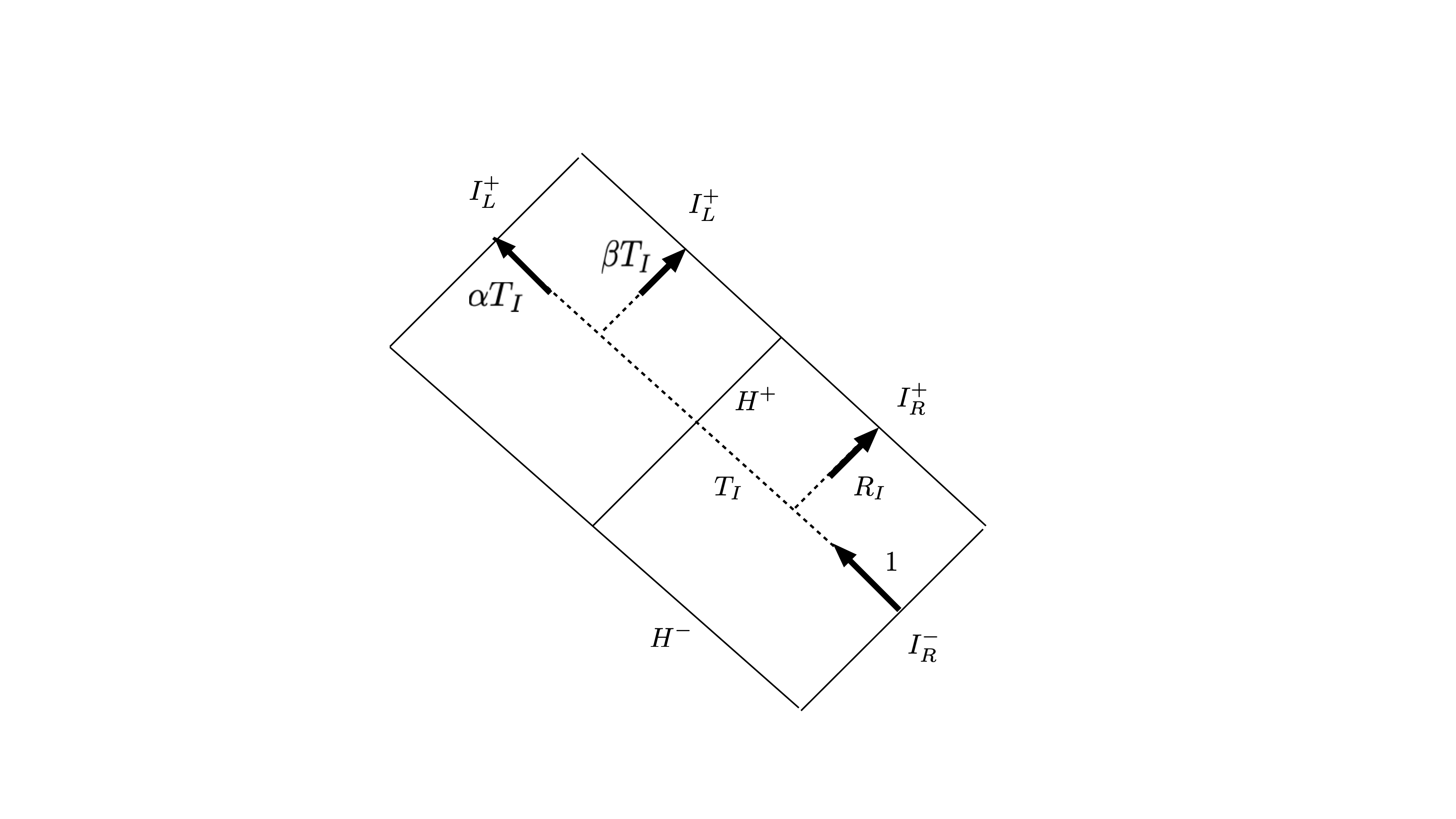}
\caption{Scattering amplitudes of the $\phi_I$ mode.}
\label{figtre}
\end{figure} 

\begin{figure}[h]
\centering 
\includegraphics[width=4.5in] {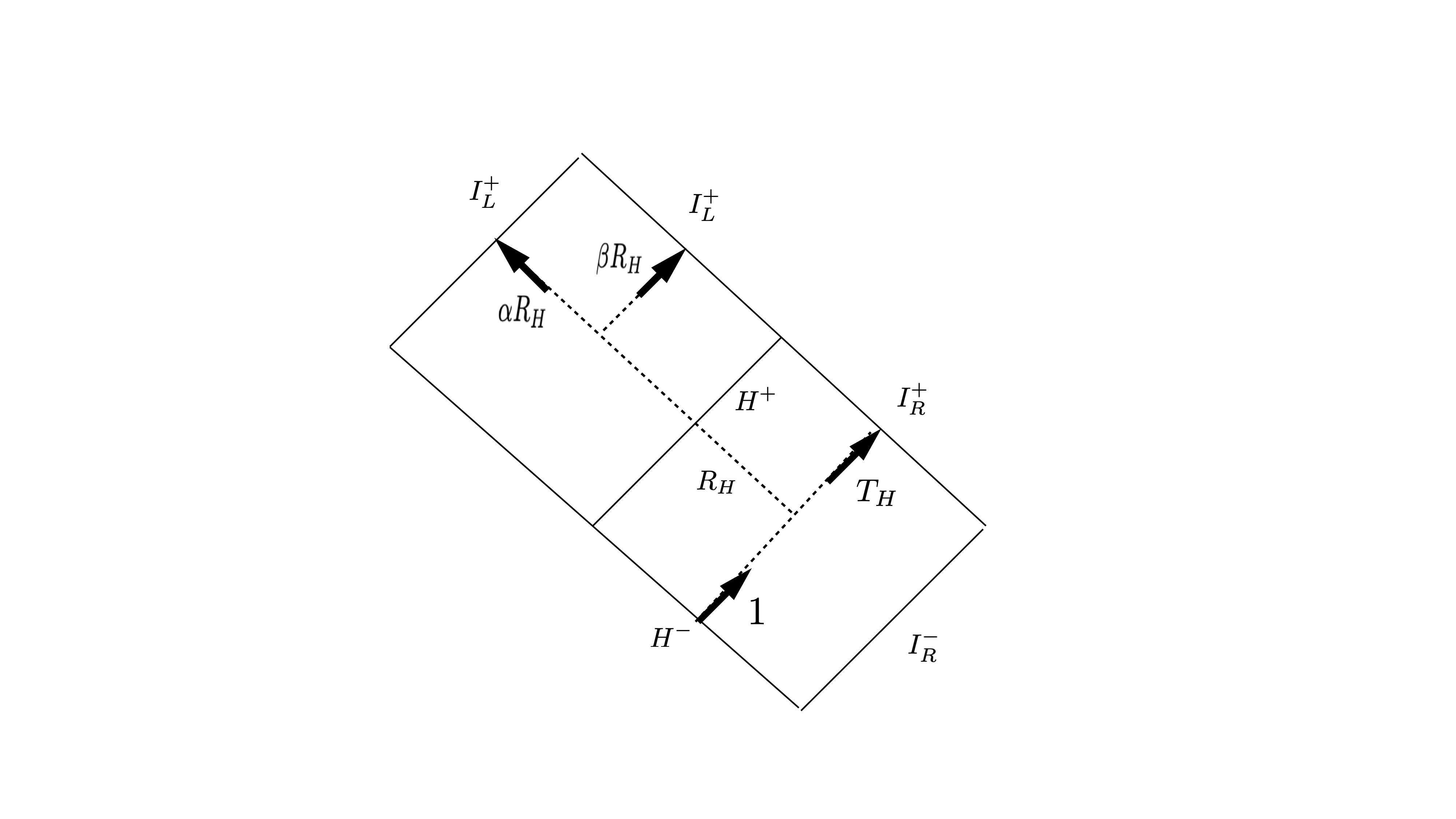}
\caption{Scattering amplitudes of the  $ \phi_H^R$ mode.}
\label{figquattro}
\end{figure} 

\begin{figure}[h]
\centering 
\includegraphics[width=4.5in] {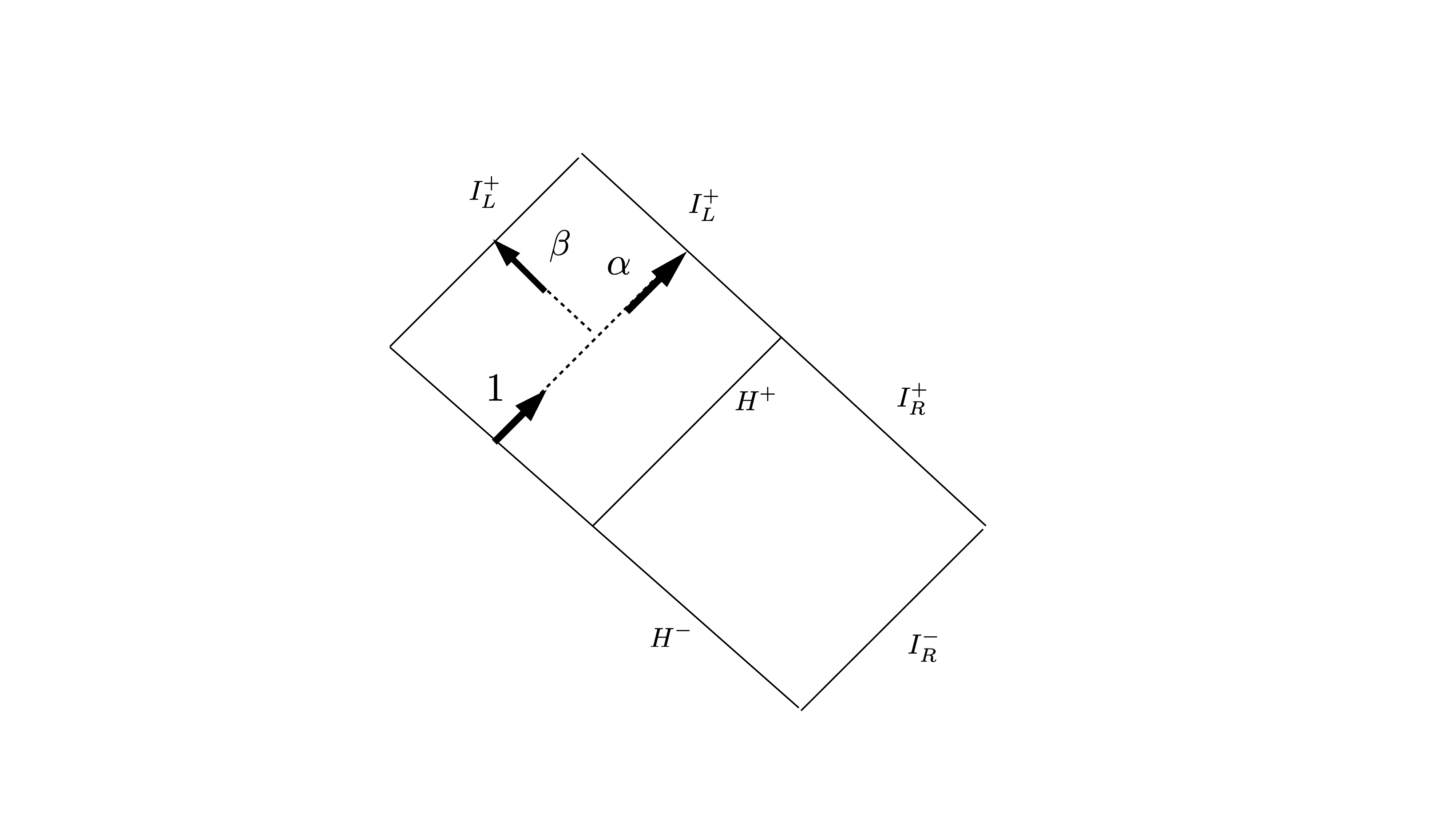}
\caption{Scattering amplitudes of the $\phi_H^L$ mode.}
\label{figcinque}
\end{figure} 

 \section{Extracting approximate analytic expressions\\ for the scattering coefficients}
 
 In this section we will obtain approximate expressions for  the scattering coefficients presented in Appendix A that will allow us, in the next section, to obtain the first analytical results for the main peaks of the density correlator. 
 We introduce two small parameters $\epsilon=1-\frac{v_0}{c_R}, \epsilon'=\frac{v_0}{c_L}-1$. We will expand all of the scattering coefficients and keep only the leading order terms in $\epsilon$ and $\epsilon'$. 
 
 The case $\epsilon=0\ ,\epsilon'=0 $ corresponds to a constant sound speed $c_R$ for $x >0$ and a constant sound speed $c_L$ for $x < 0$ 
 as can be seen in Eqs. (\ref{c-x-gt-0}) and (\ref{c-x-lt-0}) since the prefactors of the exponentials vanish.  The effective potential also vanishes, so in this limit there is no backscattering of the modes.
 
% This is nicely shown by expanding the exterior profile (\ref{c-x-gt-0}) at the leading order in $\epsilon$ 
%  \be \label{cext1} c(x)^{ext}= \frac{c_R}{\sqrt{1+ \frac{\epsilon c_R(c_R+v_0)}{|v|} e^{-\frac{2c_Rkx}{v_0(c_R+v_0)}\frac{1}{\epsilon}}}}\sim c(x)^{ext1}_{app}=c_R ( 1-\epsilon v_0e^{-kx(\frac{1}{v_0\epsilon}+\frac{1}{2})} )\ . \  \ee
%$c^{ext}_{app}$ satisfies the asymptotic limits, i.e. \be c^{ext1}_{app}(\infty)=c_R\ , \ \ c^{ext1}_{app}(0)=c_R(1-\epsilon v_0)= v_0\ . \ee
%We can also compute, from this expression, the effective potential (\ref{ep}, \ref{V})
%%\be \label{veff} V_{eff}= -\frac{(c^2-v^2)}{c}\left[ \frac{(c^2-v^2)}{c^2} \frac{c''}{2} + \frac{(5v^2-c^2)}{4c^3}c'^2\right] \ee
%we have, inserting $c^{ext1}_{app}$,
%\be \label{veff1} V_{eff} \sim 2k^2|v|^3\epsilon\  e^{-kx( \frac{1}{v_0\epsilon}+\frac{1}{2})} (1- e^{-kx (\frac{1}{v_0\epsilon}+\frac{1}{2})}) (1-2 e^{-kx(\frac{1}{v_0\epsilon}+\frac{1}{2})} ) \ . \ee
%This potential vanishes at the horizon, at $x_0=\frac{2v_0\epsilon}{(2+v_0\epsilon)k}\ln 2$ and at infinity. It has a well for $0<x<x_0$ and a barrier for $x>x_0$. When $\epsilon=0$ $V_{eff}=0$, i.e. we have no backscattering. 
% Similar results are valid for the interior profile $c(x)^{int}$ (\ref{c-x-lt-0}) in terms of the expansion parameter $\epsilon'$.
 
 We write $R_H$, from (\ref{rh}), as
 \bea \label{rha} 
 R_H &=& \frac{\Gamma(\frac{i\omega}{\kappa})}{\Gamma(-\frac{i\omega}{\kappa})}\frac{ \Gamma(-\frac{i\omega}{\kappa}+\frac{i\omega\epsilon}{2\kappa})\Gamma(1 -\frac{i\omega}{\kappa}+\frac{i\omega\epsilon}{2\kappa}) }
{ \Gamma(\frac{i\omega\epsilon}{2\kappa})\Gamma(1+ \frac{i\omega\epsilon}{2\kappa}) } \nonumber \\ &\sim& \frac{\epsilon \pi \omega}{2\kappa\sinh(\frac{\pi\omega}{\kappa})}\left( 1+\epsilon \left( -\frac{1}{2} +i\frac{\omega}{\kappa}\Big(\psi\Big(\frac{-i\omega}{\kappa}\Big)+\gamma\Big) \right) \right)\nonumber \\  &=& \frac{\epsilon \pi \omega}{2\kappa\sinh(\frac{\pi\omega}{\kappa})}\left( 1+ \frac{\epsilon  \pi\omega}{2\kappa}\coth \frac{\pi\omega}{\kappa} + \frac{i\epsilon \omega^3}{\kappa^3} \sum_{n=1}^\infty \frac{1}{n(n^2+\frac{\omega^2}{\kappa^2})} \right)
\label{rha} \ ,
\eea
where in the last line we have used \be Re \psi(\frac{-i\omega}{\kappa})=-\gamma+\frac{\omega^2}{\kappa^2} \sum_{n=1}^\infty \frac{1}{n(n^2+\frac{\omega^2}{\kappa^2})},\ Im \psi(\frac{-i\omega}{\kappa})= -\frac{\kappa}{2\omega}-\frac{\pi}{2}\coth\frac{\pi\omega}{\kappa}\ .\ee Consistently, $R_H$ vanishes when $\epsilon=0$ . Our result (\ref{rha}), for a small nonzero $\epsilon$, takes into account backscattering effects at leading order. For $T_H$ in eq. (\ref{th}), the leading order term is $1$  and the first nonzero corrections are of order 
 $\epsilon^2$
 \be \label{tha} T_H= \sqrt{\frac{v_0}{c_R}}\frac{\Gamma(-\frac{i\omega}{\kappa}+\frac{i\omega \epsilon}{2\kappa})\Gamma(1-\frac{i\omega}{\kappa}+\frac{i\omega \epsilon}{2\kappa})}{\Gamma(1-\frac{i\omega}{\kappa}+\frac{i\omega \epsilon}{\kappa})\Gamma(-\frac{i\omega}{\kappa})}\sim 1+\epsilon^2 (a+ib)\ , 
 \ee
 where $a+ib=\frac{1}{8}+\frac{\omega^2}{4\kappa^2}\psi'(-\frac{i\omega}{\kappa})$.
 The constant $a$ can also be determined by imposing unitarity $|R_H|^2+|T_H|^2\sim (\frac{\epsilon \pi \omega}{2\kappa\sinh(\frac{\pi\omega}{\kappa})})^2+ 1+2\epsilon^2 a=1$ and we get
 \be\label{coa}
 a=-\frac{\pi^2\omega^2}{8\kappa^2\sinh^2(\frac{\pi\omega}{\kappa})}\ , b=\frac{\omega^2}{4\kappa^2}Im \psi'(-\frac{i\omega}{\kappa})=\frac{\omega^3}{2\kappa^3}\sum_{n=1}^\infty \frac{n}{(n^2+\frac{\omega^2}{\kappa^2})^2} \ , \ee
 where we have used the formula ${\rm Im} \psi'(iy)=-2y\sum_{i=1}^\infty \frac{n}{(n^2+y^2)^2}$.
 By performing the same expansions for  $R_I,T_I$ in eqs. (\ref{ri}, \ref{ti}) we get
 \bea\label{rti}
 R_I&\sim&-\frac{\epsilon \pi \omega}{2\kappa\sinh(\frac{\pi\omega}{\kappa})}\left( 1+ \frac{\epsilon  \pi\omega}{2\kappa}\coth \frac{\pi\omega}{\kappa} - \frac{i\epsilon \omega^3}{\kappa^3} \sum_{n=1}^\infty \frac{1}{n(n^2+\frac{\omega^2}{\kappa^2})} \right)
 \ , 
 \\ T_I &\sim& 1+\epsilon^2(a+ib) \;, \nonumber 
 \eea 
 and for $\alpha,\beta$ in eqs. (\ref{al}, \ref{be}), we find %, (\ref{tabe})
 \bea \label{abea}
 \alpha &\sim& 1-\epsilon'^2(a+ib)\ , \\ \beta &\sim& \frac{\epsilon' \pi \omega}{2\kappa\sinh(\frac{\pi\omega}{\kappa})}\left( 1- \frac{\epsilon'  \pi\omega}{2\kappa}\coth \frac{\pi\omega}{\kappa} + \frac{i\epsilon' \omega^3}{\kappa^3} \sum_{n=1}^\infty \frac{1}{n(n^2+\frac{\omega^2}{\kappa^2})} \right)
\ .  \nonumber \eea
 %\label{tabea} \ee
% \tilde \alpha\sim 1-\epsilon'^2(a-ib)\ , \tilde \beta \sim \frac{\epsilon' \pi |v|\omega}{2\kappa\sinh(\frac{\pi\omega}{\kappa})}\ . \eea 
 
  \section{Identifying the three relevant correlators\\ from the two-point function}
 
 Referring to the general expressions (\ref{tpf}) - (\ref{Jdef}) for the two-point function,
 we will identify here the expressions for the three relevant correlators \cite{paper2013}. Two of them are found by considering one point outside and the other point inside the horizon, and the remaining one when both points are inside the horizon. 
 %Let us consider first the two-point function of the phase fluctuation $\hat\phi$. 
 %In the external region it takes the form
%\be \label{ff}
%\langle \{ \hat\phi  (t,x)\hat \phi (t',x') \} \rangle  = I+J\ , \ee
%with
%\bea \label{i} I&=& \int_0^\infty d\omega  \coth (\frac{\pi\omega}{\kappa}) \left( \varphi_H(x)\varphi_H^*(x')e^{-i\omega(t_S-t_S')}+c.c. \right) \ .   \\
%J&=&\int_0^\infty d\omega \left( \varphi_I(x)\varphi_I^*(x')e^{-i\omega(t_S-t_S')}+c.c. \right) \label{j}  %\eea
%and the modes $\chi_H$ and $\chi_I$ are defined in eqs. (2.8), (2.9) of the draft ($\kappa=k|v|$ is the %surface gravity, see eq. (3.5) of the draft).
For one point outside and one inside we have, in general,
\bea \label{df} && \langle \{ \hat\phi  (T,x)\hat \phi (T',x') \} \rangle  = \int_0^\infty \frac{d\omega}{\sinh(\frac{\pi\omega}{ \kappa})}\left( \varphi_H(x)\varphi^{int}(x')e^{-i\omega(t-t')}
+c.c. + (x\leftrightarrow x') \right) + \\ 
 &&\int_0^\infty d\omega \left[ \left( \varphi_I(x)T_I^*\varphi^{int\ *}(x') +\coth (\frac{\pi\omega}{\kappa}) \varphi_H(x) R_H^* \varphi^{int\ *}(x') \right) e^{-i\omega(t-t')}
 +c.c. + (x\leftrightarrow x')\right], \nonumber  \eea
%where $\chi_2^{int}$ and $\chi_1^{int}$ have been defined in eqs. (3.24) and (3.21)  of the draft. 
and for both points inside 
\bea &&  \langle \{ \hat\phi  (T,x)\hat \phi (T',x') \} \rangle=  \int_0^\infty d\omega  \coth (\frac{\pi\omega}{\kappa}) 
\left( \varphi^{int*}(x)\varphi^{int}(x')e^{-i\omega(t-t')}+c.c. \right) \nonumber \\  
&& + \int_0^\infty d\omega\Big[ \Big( |T_I|^2\varphi^{int}(x)\varphi^{int\ *}(x')+ \coth (\frac{\pi\omega}{\kappa}) |R_H|^2 \varphi^{int}(x)\varphi^{int\ *}(x')\nonumber \\ && + 
\frac{1}{\sinh(\frac{\pi\omega}{ \kappa})} \varphi^{int*}(x)R_H^*\varphi^{int\ *}(x')\Big)e^{-i\omega(t-t')}+c.c. \Big]\ . \label{dau}
\eea
%Note that, with the respect to the standard expressions, the 2nd line of (\ref{df}) and 2nd and 3rd lines of (\ref{dd}) arise by considering, explicitly, the continuation of the $\chi_I$ and $\chi_H$ modes inside the horizon, which in the draft, eqs. (2.8) and (2.9),  have been defined only outside. 
By considering the chosen points far outside and  inside the horizon, we use the asymptotic expansions of the modes in terms of the scattering coefficients and use the null $u,v$ coordinates (respectively, outgoing and ingoing Eddington-Finkelstein coordinates)
%(introduced in section 1) 
defined in terms of $(t,z)$ and laboratory coordinates $(T,x)$ asymptotically as
\bea v_{R,L} &=& t + \frac{z}{c_{R,L}}= T+\frac{x}{c_{R,L}+v_0}\ , \\ u_{R,L}&=& t - \frac{z}{c_{R,L}}=T-\frac{x}{c_{R,L}-v_0}\ .\eea Referring to (\ref{df}), the relevant correlators are those between $u_R-u'_L, u_R'-u_L, u_R-v_L',v_L-u_R'$
\bea && \langle \hat \phi \hat \phi \rangle _{ext-int}= \frac{m}{n}\frac{\sqrt{c_Rc_L}}{4\pi}\Big[ \int_0^\infty \frac{d\omega}{\omega\sinh(\frac{\pi\omega}{\kappa})}\Big(T_H\alpha e^{-i\omega\Delta u}+T_H\beta e^{-i\omega(u-v')} \Big)+ \nonumber \\ 
&& \int_0^\infty \frac{d\omega}{\omega} \Big( 
%T_I^*\alpha^*e^{-i\omega\Delta v} + T_I^*\beta^*e^{-i\omega(v-u')}+
R_IT_I^*\alpha^*e^{-i\omega(u-v')}+
R_IT_I^*\beta^*e^{-i\omega\Delta u}\Big)+\nonumber \\
&&\int_0^\infty \frac{d\omega}{\omega}\frac{\cosh(\frac{\pi\omega}{\kappa})}{\sinh(\frac{\pi\omega}{\kappa})}\Big( T_HR_H^*\alpha^*e^{-i\omega(u-v')}+T_HR_H^*\beta^*e^{-i\omega\Delta u}\Big)
+c.c.+(x\leftrightarrow x') \Big]\  . \label{coco}
\eea
We insert in the above the approximate expressions for the scattering coefficients derived in the previous section and, by considering the point $x$ outside the horizon and $x'$ inside, we separate the Hawking quanta - partner (HP)  correlator ($\Delta u=u_R-u_L'$)
\bea \langle\hat\phi\hat\phi\rangle_{H-P} &=&\frac{m}{n}\frac{\sqrt{c_Rc_L}}{2\pi} \int_\lambda^\infty \frac{d\omega}{\omega} \Big[ \frac{ 1+ \frac{(\epsilon'^2-\epsilon^2)A\omega^2}{2\sinh^2(\frac{\pi\omega}{\kappa})}  }  {\sinh(\frac{\pi\omega}{\kappa})}  \cos(\omega\Delta u)
+  b(\epsilon^2-\epsilon'^2)\frac{\sin(\omega\Delta u)}{\sinh(\frac{\pi\omega}{\kappa})} 
\nonumber \\ &-& \frac{\epsilon\epsilon'A\omega^2}{\sinh^2(\frac{\pi\omega}{\kappa})}
\left(1- \frac{\cosh(\frac{\pi\omega}{\kappa})}{\sinh(\frac{\pi\omega}{\kappa})}\right)
\cos(\omega\Delta u) \Big]\ ,   \label{hp}
\eea
where $ A=\frac{\pi^2 }{4\kappa^2}\ $,
and the $(u_R,v'_L)$ correlator
\be  \langle \hat\phi\hat\phi \rangle_{u_R,v'_L}^{int-ext}=
% \frac{m}{n}\frac{\sqrt{c_Rc_L}}{4\pi}\int_0^\infty \frac{d\omega}{\omega} \Big( \frac{T_H\beta}{\sinh(\frac{\pi\omega}{\kappa})}
%+R_IT_I^*\alpha^*+
%\nonumber \\ && T_HR_H^*\alpha^*\coth(\frac{\pi\omega}{\kappa})\Big) e^{-i\omega(u-v')}+c.c.  \nonumber \\
%\be  \langle \hat\phi\hat\phi \rangle_{u,v'}^{int-ext}
 \frac{m}{n}\frac{\sqrt{c_Rc_L}}{4\pi} \frac{\pi }{\kappa}\int_0^\infty d\omega \left( \frac{\epsilon'}{\sinh^2(\frac{\pi\omega}{\kappa})}
-\frac{\epsilon}{\sinh(\frac{\pi\omega}{\kappa})}(1- \coth(\frac{\pi\omega}{\kappa}))\right) \cos(\omega(u_R-v'_L))\ .  \label{urvl}
\ee
Finally, if we consider $x,x'$ in the far inside region from (\ref{dau}) we get an expression depending on the scattering coefficients and $u_L-v_L', v_L-u_L'$. By considering again the expansions  for the various scattering coefficients derived in the previous section,  the $(u_L,v_L')$ correlator takes the form 
\be   \langle\hat\phi\hat\phi\rangle_{u_L,v'_L} =\frac{m}{n}\frac{c_L}{4\pi} \int_0^\infty d\omega \frac{\pi }{\kappa}  \Big[ \frac{\epsilon'\cosh(\frac{\pi\omega}{\kappa})}{\sinh^2(\frac{\pi\omega}{\kappa}) }+\frac{\epsilon'}{\sinh(\frac{\pi\omega}{\kappa})}+ \frac{\epsilon}{\sinh^2(\frac{\pi\omega}{\kappa})}\Big]\cos\omega(u_L-v_L')\ . \label{ulvl}  \ee

It is clear from the obtained expressions that, while the correlator in (\ref{hp}) is there even for $\epsilon,\epsilon' \to 0$ (in this limit it formally coincides with the 2D expression, without backscattering), the other two signals disappear in the same limit (their presence is due to backscattering).

 \section{Analytical results for the peaks locus and heights of the three nontrivial correlators in the density correlator}
 
 Our purpose is to get analytic expressions for the heights of the relevant peaks  of the density-density correlation function (\ref{G2-eq}) derived from (\ref{hp}), (\ref{urvl}), (\ref{ulvl}). 
%\be \label{dd} G_2(x,x')=\lim_{t\to t'} \frac{n^2}{2m^2c^2(x)c^2(x')}\left( \partial_t\partial_{t'}-|v|(\partial_t\partial_{x'}+\partial_{t'}\partial_x) + v^2\partial_x\partial_{x'} \right) \langle \{ \hat\phi  (t,x)\hat \phi (t',x') \} \rangle \ . \ee
 %thus improving the free field results. 
 
 For the HP correlator (\ref{hp}), by performing the derivatives in (\ref{G2-eq}) we get\footnote{Here we have inserted in the results the healing length $\xi=\frac{\hbar}{mc}$, as in \cite{paper1}.} 
 \bea &&\frac{G_2^{HP}}{n^2}= - \frac{1}{4\pi} \frac{\xi_L\xi_R}{c_Lc_R}\frac{1}{\sqrt{(n\xi_L)(n\xi_R)}}\frac{c_Rc_L}{(c_R - v_0)(v_0 -c_L)} 
\Big[ \int_0^\infty d\omega \frac{ \omega\cos(\omega\Delta u)}{\sinh(\frac{\pi\omega}{\kappa})} \nonumber \\
&&+ \frac{(\epsilon'^2-\epsilon^2)\pi^2}{8\kappa^2}\int_0^\infty d\omega \frac{ \omega^3\cos(\omega\Delta u)}{\sinh^3(\frac{\pi\omega}{\kappa})}
-\frac{(\epsilon^2-\epsilon'^2)}{2\kappa^3} \sum_{n=1}^\infty \int_0^\infty d\omega \frac{\omega^4 n}{(n^2 + \frac{\omega^2}{\kappa^2})^2} \frac{\sin(\omega\Delta u)}{\sinh(\frac{\pi\omega}{\kappa})} \nonumber \\ && -\frac{\epsilon\epsilon' \pi^2 }{4\kappa^2}\left( \int_0^\infty d\omega \frac{ \omega^3\cos(\omega\Delta u)}{\sinh^2(\frac{\pi\omega}{\kappa})}
- \int_0^\infty d\omega \frac{ \omega^3 \cosh(\frac{\pi\omega}{\kappa})\cos(\omega\Delta u)}{\sinh^3(\frac{\pi\omega}{\kappa})}\right)  \Big]_{t=t'} \ . \label{hpdc} \eea

The first integral gives $\frac{\kappa^2}{4\cosh^2(\frac{\kappa\Delta u}{2})}$ and reproduces the 2D correlator \cite{paper1}. The other integrals are given in Appendix B.

Combining all of the integrals, we find 
\bea &&\frac{G_2^{HP}}{n^2}= - \frac{1}{4\pi} \frac{\xi_L\xi_R}{c_Lc_R}\frac{1}{\sqrt{(n\xi_L)(n\xi_R)}}\frac{c_Rc_L}{(c_R - v_0)(v_0-c_L)} 
\Big[ \frac{\kappa^2}{4\cosh^2(\frac{\kappa\Delta u}{2})} + \nonumber \\ &&
 \frac{(\epsilon'^2-\epsilon^2)\kappa^4}{128\cosh^4(\frac{\kappa\Delta u}{2})}\left( (\Delta u^2 + \frac{(6+\pi^2)}{\kappa^2} ) 
\cosh(\kappa \Delta u) +\frac{6}{\kappa^2} -2(\Delta u^2 +\frac{\pi^2}{\kappa^2}) -6\frac{\Delta u}{\kappa}\sinh(\kappa\Delta u) \right)\ \ \ \ \nonumber \\ &&
-\frac{(\epsilon^2-\epsilon'^2)}{2\kappa^3} \sum_{n=1}^\infty \int_0^\infty d\omega \frac{\omega^4 n}{(n^2 + \frac{\omega^2}{\kappa^2})^2} \frac{\sin(\omega\Delta u)}{\sinh(\frac{\pi\omega}{\kappa})} \nonumber \\ && +\frac{\epsilon\epsilon' \kappa^2}{16\pi^2}\left( 3Re\psi''(-\frac{i\kappa\Delta u}{2\pi})-\frac{\Delta u}{2}\frac{\kappa}{\pi}Im\psi'''(-\frac{i\kappa\Delta u}{2\pi})\right)  \nonumber \\ &&
- \frac{\epsilon\epsilon' \kappa^4}{64\sinh^4(\frac{\kappa\Delta u}{2})}\Big[ (\frac{6}{\kappa^2}+\Delta u^2)\cosh(\kappa \Delta u) -\frac{6}{\kappa^2}+2\Delta u^2 -6\frac{\Delta u}{\kappa}\sinh(\kappa\Delta u) \Big]\Big]_{t=t'} \ . \label{aaa} \eea

To look for the location of the peak, we consider the expansion of the above expression inside the square brackets for small $\Delta u$. The first term gives $\frac{\kappa^2}{4}(1-(\frac{\kappa\Delta u}{2})^2)$, the terms in the second, fourth and fifth lines have leading terms that are constants and are second order in the expansion parameters. The sum term in the third line gives a linear order term in $B\Delta u$, where $B=-\frac{46.4224(\epsilon^2-\epsilon'^2)  \kappa^3}{2\pi^6}$. Putting it all together, we see that the location of the peak is at $\Delta u= \frac{8 B}{\kappa^4}$, which is (slightly) shifted with respect to the location it has in the 2D conformal theory ($\Delta u=0$, which at $t=t'$ corresponds to $\frac{x}{c_R-v_0}+\frac{x'}{v_0-c_L}=0$ \cite{paper1}). 

Since the peak location is second order in the expansion parameters, to evaluate the height of the peak it is enough to substitute $\Delta u=0$ in the expression (\ref{aaa}). 
%Alternatively, we can put $\Delta u=0$ directly in (\ref{hpdc}) (the two procedures give the same result) and using the integrals $\int_0^\infty d\omega \frac{\omega}{\sinh(\frac{\pi\omega}{\kappa})}=\frac{\kappa^2}{4}, \int_0^\infty d\omega \frac{\omega^3}{\sinh(\frac{\pi\omega}{\kappa})^3}=\frac{\kappa^4}{16\pi^2}(12-\pi^2), \int_0^\infty d\omega \frac{\omega^3}{\sinh(\frac{\pi\omega}{\kappa})^2}=\frac{3\kappa^4}{2\pi^4}\zeta(3), \int_0^\infty d\omega\omega^3 \frac{\cosh(\frac{\pi\omega}{\kappa}) }{\sinh(\frac{\pi\omega}{\kappa})^3}=\frac{\kappa^4}{4\pi^2}$. 
We get
\bea \frac{G_2}{n^2}^{HP}|_{peak} &=& - \frac{\kappa^2}{16\pi} \frac{\xi_L\xi_R}{c_Lc_R}\frac{1}{\sqrt{(n\xi_L)(n\xi_R)}}\frac{c_Rc_L}{(c_R - v_0)(v_0 - c_L)} \times \nonumber \\
&& \Big[ 1 + \frac{(12-\pi^2)(\epsilon'^2-\epsilon^2)}{32}- \frac{3\epsilon\epsilon'  \zeta(3)}{2\pi^2} + \frac{\epsilon\epsilon' }{4} \big]\ .\eea

 %By fixing $c_L=1/2, v_0=\frac{3}{4}, c_R=1$, 
%we have $\epsilon=\frac{1}{4},\ \epsilon'=\frac{1}{2}$ (both $<1$) and ($\zeta(3)=1.20206$)

%{\bf{ 
We can compare our analytical results with numerical studies of the peaks structure present in the literature \cite{paper-renaud, paper2013, paper2}, where a similar sound velocity profile is considered with the following values: $c_L=\frac{1}{2}, v_0=\frac{3}{4}, c_R=1$, expressed in units of $c_R$. Note that the corresponding values of $\epsilon=\frac{1}{4}, \epsilon'=\frac{1}{2}$ are not much smaller than $1$ and so our (leading) second order expansion in these parameters can be just expected to give qualitative agreement with the numerics as we will show. 
%WE CAN COMPARE OUR ANALYTICAL RESULTS WITH NUMERICAL STUDIES OF THE PEAKS STRUCTURE PRESENT IN THE LITERATURE [15,18,5] WHERE A SIMILAR SOUND VELOCITY PROFILE IS CONSIDERED WITH THE FOLLOWING VALUES: $C_L=1/2$ , $V_0 = 3/4$ AND $C_R =1$ EXPRESSED IN UNITS OF $C_R$. NOTE THAT THE CORRESPONDING VALUES OF $\epsilon= 1/4$ AND $\epsilon'= 1/2$ ARE NOT MUCH SMALLER THAN 1 AND SO OUR SECOND ORDER EXPANSION IN THESE PARAMETERS CAN BE JUST EXPECTED TO GIVE QUALITATIVE AGREEMENT WITH THE NUMERICS AS WE WILL SHOW.
Using $\zeta(3)=1.20206$\  we have 
%}}

  \be\label{are} \frac{(12-\pi^2)(\epsilon'^2-\epsilon^2)}{32}- \frac{3\epsilon\epsilon'  \zeta(3)}{2\pi^2} + \frac{\epsilon\epsilon' }{4} = 0.0209  \ .\ee 
Our (leading order in $\epsilon, \epsilon'$) relative correction is $2 \times 10^{-2}$, and (slightly) increases the value of the Hawking peak with respect to its unperturbed value ($\epsilon=\epsilon'=0$, which is the one given by the 2D theory). 

%{\bf{
Considering the numerics, we see that  the shift and the increase in the height of the peak appear clearly in the right plots of Fig. (12) of \cite{paper-renaud}:  at fixed $x'$ (for which $\Delta u=-\frac{\Delta x}{(c_R-v_0)}$) we find, analytically, that the shift to the peak location with respect to that predicted by the 2D conformal theory ($\Delta u=0$)  is negative ($\Delta x<0$), in agreement with the observed shift to the left of the orange numerical (with the same profile considered in this paper) dotted line compared to the 2D analytical solid orange one. %(the numerical shift is $\Delta x=..$). 
This figure also confirms the slight increase of the (absolute value of) the peak in the dotted curve with respect to that in the solid one. The difference between the numerical height and the 2D height (divided by the  2D height) is given by 0.018, the analytical value is given in (\ref{are}), and the relative difference is
\be \label{rdhp} \frac{G_{2\ numeric}^{HP}-G_{2\ analytic}^{HP}}{G_{2\ numeric}^{HP}}|_{peak} =
 -0.003\ .\ee  The qualitative agreement (order of magnitude and sign of the displacement) with our analtical predictions is quite good.
 %}}
%THE LOCATION OF THE PEAK IS AT.... 
%CONSIDERING THE NUMERICS, WE SEE THAT THE SHIFT AND THE INCREASE IN HEIGHT OF THE PEAK APPEAR CLEARLY IN THE RIGHT PLOT OF FIG.  12 OF [18]: AT FIXED X’ (FOR WHICH  $\Delta U=-\Delta X/(C_R-V_0)$ ) WE FIND, ANALYTICALLY, THAT $\Delta X<0$, IN AGREEMENT WITH THE OBSERVED SHIFT TO THE LEFT OF THE ORANGE NUMERICAL (WITH THE SAME PROFILE CONSIDERED IN THIS PAPER) DOTTED LINE VS THE 2D ANALYTICAL SOLID ORANGE ONE. THIS FIGURE ALSO CONFIRMS THE SLIGHT INCREASE OF THE (ABSOLUTE VALUE OF THE) PEAK IN THE DOTTED CURVED WITH RESPECT TO THAT IN THE SOLID ONE.THE VALUE OF THE SHIFT AND HEIGHT INCREASE COMING FROM THE NUMERICAL ANALYSIS ARE........ RESPECTIVELY.
%THE QUALITATIVE AGREEMENT ( ORDER OF MAGNITUDE AND SIGN OF THE DISPLACEMENT) WITH OUR ANALYTIC PREDICTIONS IS QUITE GOOD.

For the $(u_R,v_L')$ correlator (\ref{urvl}) at leading order we have
\bea
&&\frac{G^{(2)}_{u_R,v_L'}}{n^2}=\frac{1}{4\pi}\frac{\xi_L\xi_R}{c_Lc_R}\frac{1}{\sqrt{(n\xi_L)(n\xi_R)}}\frac{c_Rc_L}{(c_R - v_0)(c_L+v_0 )} \frac{\pi }{2\kappa}\times \nonumber \\
&& \int_0^\infty d\omega \left( \frac{\epsilon'\ \omega^2}{\sinh^2(\frac{\pi\omega}{\kappa})}-\frac{\epsilon\ \omega^2}{\sinh(\frac{\pi\omega}{\kappa})}
+\frac{\epsilon\ \omega^2 \cosh (\frac{\pi\omega}{\kappa})}{\sinh^2(\frac{\pi\omega}{\kappa})} \right) \cos(\omega(u_R-v'_L))|_{t=t'}\ . \label{scll} \eea
Using the results for the integrals in Appendix B we have
 \bea
&&\frac{G^{(2)int-ext}_{u_R,v_L'}}{n^2}=\frac{1}{4\pi}\frac{\xi_L\xi_R}{c_Lc_R}\frac{1}{\sqrt{(n\xi_L)(n\xi_R)}}\frac{c_Rc_L}{(c_R - v_0)(c_L+v_0 )} \times \nonumber \\
&& \Big[ \frac{\epsilon'  \kappa^2}{8} \frac{\left( -2+(u-v')\kappa \coth(\frac{\kappa(u-v')}{2})\right)}{\sinh^2(\frac{\kappa(u-v')}{2})} + \frac{\epsilon  \kappa^2}{8\pi^2}Re \psi'' (-\frac{i\kappa(u-v')}{2\pi}+\frac{1}{2})\nonumber \\ &&  -\frac{\epsilon  \kappa^2}{8}\frac{\left( -2+(u-v')\kappa \tanh(\frac{\kappa(u-v')}{2})\right)}{\cosh^2(\frac{\kappa(u-v')}{2})}   \Big]_{t=t'} \label{spll} \  .
\eea 
Our leading order $\epsilon,\epsilon'$ expansion (\ref{spll})  has a peak at $u_R-v'_L=0$ (at $T=T'$, $\frac{x}{c_R-v_0}+\frac{x'}{c_L+v_0}=0$). We can see this by expanding it for small $u_R-v_L'$: we get an expression of the type $A+B(u_R-v_L')^2$, for some constants $A,B$ (at first order in $\epsilon, \epsilon'$). Such an expression is extremal for $u_R-v_L'=0$. However, we can have a more refined estimation for the peak locus by considering the second order terms of $R_H, R_I, \beta$ given in (\ref{rha}), (\ref{rti}), (\ref{abea}). 
%In particular, their imaginary parts provide 
%a linear (small) $u_R-v_L'$ term of second order in $\epsilon^2,\epsilon'^2$ in the correlator. 
Specifically, we find that the relevant terms in the small $u_R-v_L'$ expansion of $\frac{G^{(2)int-ext}_{u_R,v_L'}}{n^2}$ from (\ref{urvl}) expanded up to 2nd order are proportional to $-\frac{(u_R-v_L')^2}{2}(\frac{\pi^4\epsilon'}{30}-48.2171\epsilon +\frac{\epsilon \pi^4}{2})+ \frac{(u_R-v_L')}{\kappa}(1.51525\epsilon'^2+48.5833\epsilon^2-48.7045\epsilon^2)$. Such an expression is extremal for $u_R-v_L'$ given by a first order expression in the perturbation parameters.

As for the height of the peak, at leading order it is enough to substitute  $u_R-v'_L=0$ in (\ref{spll})
%or, alternatively, substitute it directly in (\ref{scll}) and using the integrals $\int_0^\infty d\omega \frac{\omega^2}{\sinh^2(\frac{\pi\omega}{\kappa})}=\frac{\kappa^3}{6\pi},  \int_0^\infty d\omega \frac{\omega^2}{\sinh(\frac{\pi\omega}{\kappa})}=\frac{7\zeta(3)\kappa^3}{2\pi^3}, \int_0^\infty d\omega \frac{\omega^2 \cosh(\frac{\pi\omega}{\kappa})}{\sinh^2(\frac{\pi\omega}{\kappa})}=\frac{\kappa^3}{2\pi}$. 
and get
\be \frac{G^{(2)}_{u_R,v'_L}}{n^2}|_{peak} =\frac{1}{4\pi}\frac{\xi_L\xi_R}{c_Lc_R}\frac{1}{\sqrt{(n\xi_L)(n\xi_R)}}\frac{c_Rc_L}{(c_R - |v|)(c_L+|v| )} \frac{\kappa^2}{4} 
\big[\frac{\epsilon' }{3}  -\frac{\epsilon }{2\pi^2}(16.8288) +\epsilon  \big]. \ee
%where $\frac{\epsilon'|v|}{3}  -\frac{\epsilon|v|}{2\pi^2}(16.8288) +\epsilon |v|=
Then the ratio of the height of this peak to that of the main peak is
\be \frac{G^{(2)}_{u_R,v'_L}|_{peak}}{|G_2^{HP}|_{peak}|}= \frac{(v_0-c_L)}{(v_0+c_L)}\ \big[\frac{\epsilon' }{3}  -\frac{\epsilon }{2\pi^2}(16.8288) +\epsilon  \big]=0.0407055 \ .
\ee 
Thus, this peak is $25$ times smaller than the H-P peak. Note that the ratio of the two prefactors in $G_2$ contributes $0.2$, a further $0.2$ coming from the terms in the square parenthesis. A full numerical analysis based on the profile used in this paper provides, for this rate, the value $0.0302$ \cite{paper-renaud}.

Finally, for the $(u_L,v'_L)$ correlator (\ref{ulvl}) we have 
\bea
&&\frac{G^{(2)}_{u_L,v_L'}}{n^2}=-\frac{1}{4\pi}\frac{\xi_L^2}{c_L^2}\frac{1}{(n\xi_L)}\frac{c_L^2}{(v_0-c_L)(c_L+v_0 )} \frac{\pi }{2\kappa}\times \nonumber \\
&& \int_0^\infty d\omega \left( \frac{\epsilon\ \omega^2}{\sinh^2(\frac{\pi\omega}{\kappa})}+\frac{\epsilon' \omega^2}{\sinh(\frac{\pi\omega}{\kappa})}
+\frac{\epsilon' \omega^2  \cosh (\frac{\pi\omega}{\kappa})}{\sinh^2(\frac{\pi\omega}{\kappa})} \right) \cos(\omega(u-v')) |_{t=t'}\  \eea
and we get
 \bea
&&\frac{G^{(2)int-int}_{u_L,v_L'}}{n^2}=-\frac{1}{4\pi}\frac{\xi_L^2}{c_L^2}\frac{1}{(n\xi_L)}\frac{c_L^2}{(v_0-c_L)(c_L+v_0 )} \times \nonumber \\
&& \Big[ \frac{\epsilon  \kappa^2}{8} \frac{\left( -2+(u-v')\kappa \coth(\frac{\kappa(u-v')}{2})\right)}{\sinh^2(\frac{\kappa(u-v')}{2})} - \frac{\epsilon'  \kappa^2}{8\pi^2}Re \psi'' (-\frac{i\kappa(u-v')}{2\pi}+\frac{1}{2})\nonumber \\ &&  -\frac{\epsilon'  \kappa^2}{8}\frac{\left( -2+(u-v')\kappa \tanh(\frac{\kappa(u-v')}{2})\right)}{\cosh^2(\frac{\kappa(u-v')}{2})}   \Big]_{t=t'} \  . \label{uaua} \eea

As for the correlator (\ref{spll}), the peak of the above expression at leading order in $\epsilon,\epsilon'$ is located at  $u_L-v'_L=0$ (at $T=T', \frac{x}{v_0-c_L}+\frac{x'}{c_L+v_0}=0$).  A similar argument to that given after (\ref{spll}) can be made here as well, showing that even in this case calculation of $\frac{G^{(2)int-int}_{u_L,v_L'}}{n^2}$ at second order shows that the peak locus is at $u_L-v_L'=O(\epsilon,\epsilon')$. 
 Thus, to determine the peak height at leading order, it is enough to substitute $u_L-v_L'=0$ in (\ref{uaua}).  We find
\be \frac{G^{(2)int-int}_{u_L,v_L'}|_{peak}}{G_2^{HP}|_{peak}}= \sqrt{\frac{c_R}{c_L}}\frac{(c_R-v_0)}{(v_0+c_L)}\ \big[\frac{\epsilon }{3}  +\frac{\epsilon' }{2\pi^2}(16.8288) +\epsilon'  \big]=0.285561 \ . \ee
This peak is $3.5$ times smaller than the Hawking peak. This value is approximately given by the ratio of the two prefactors ($0.282843$), the terms in the square parenthesis contributing a factor of $1.00961$.
A full numerical analysis using this profile provides, for this ratio, the value 0.191 \cite{paper-renaud}.

\section{Conclusions}

In an inhomogeneous BEC, phase fluctuations in the hydrodynamical approximation propagate like a massless scalar field in a curved (fictitious) space-time. Because of curvature, the modes get backscattered in their journey. By neglecting backscattering one can approximate the theory to that of a massless scalar field that is conformally invariant in two dimensions, which can be solved exactly. The corresponding density-density correlation function in the case of an acoustic BH shows a (negative) peak when one point is outside the horizon and the other inside. Including backscattering makes the picture more intriguing highlighting the presence of two further (minor) peaks. However, this richer structure has only been previously seen in numerical computations.

Here, for the first time, a complete analytical analysis in the presence of backscattering is made with a smooth profile which allows the main features of the peaks including their origin, position and height to be estimated. We have seen how the backscattering can shift the position of the peaks from the ones predicted by geometric optics. Also, the height of the main peak is slightly modified with respect to the no-backscattering 
%{\bf{
approximation. All these are in good qualitative agreement with the numerical analysis.
%}}
%APPROXIMATION. ALL THESE ARE IN GOOD QUALITATIVE AGREEMENT WITH THE NUMERICAL ANALYSIS.
As mentioned in the Introduction, at present only the main peak has been observed experimentally. The experiments do not appear to have had the accuracy required for the observation of the two minor peaks and the fine structure described here.

It was eight years between the prediction of the existence of the main peak as signature for Hawking radiation in a BEC analogue BH to its actual experimental observation. We feel confident that the features described here will be observed in a shorter time than that.

\acknowledgments

A.F. acknowledges partial financial support by the Spanish Grants PID2020-116567GB-C21, PID2023-149560NB-C21 funded by MCIN/AEI/10.13039/501100011033, and by the Severo Ochoa Excellence Grant CEX2023-001292-S. P. R. A. was supported in part by the National Science Foundation under grant No. PHY-2309186. D.P. acknowledges the University of Valencia for a doctoral grant.

\begin{appendix}
\section{DERIVATION OF THE FULL SOLUTION TO THE WAVE EQUATION FOR $\hat\phi$}
\label{appendixA}
%\section{Appendix A}

In this Appendix, we briefly describe the explicit analytic construction of the modes $\phi_I, \phi_H^R, \phi_H^L$.
As discussed in \cite{fba2016}, 
%~\cite{fba2016}, 
there are two linearly independent solutions to~\eqref{varphi-z-eq} using the general form of the profile~\eqref{c-profile-1}.  With the notation
\bes \bea y_\pm &\equiv & \frac{1}{2} (1 \mp \tanh kz) \;, \\
     \gamma_\pm &\equiv& \sqrt{A \pm B} \;,  \\
     \gamma_\mp &\equiv& \sqrt{A \mp B} \eea \ees
they are
\be \varphi_{\pm}  = y_\pm^{-i \frac{\w \gamma_\pm}{2 k}} (1 - y_\pm)^{i \frac{\w \gamma_\mp}{2 k}} F(a_\pm, b_\pm, c_\pm; y_\pm)  \;.  \ee
Here $F$ is a hypergeometric function \cite{bed, abs} and
\bea a_\pm &=& 1 - \frac{i \w}{2 k} (\gamma_\pm - \gamma_\mp)\;, \nonumber \\
     b_\pm &=& - \frac{i \w}{2 k} (\gamma_\pm - \gamma_\mp)\;, \nonumber \\
     c_\pm &=& 1 - \frac{i \w}{k} \gamma_\pm \;. \eea
For the exterior region we denote these by $\varphi^R_\pm$ and use $A_R$ and $B_R$ in the above expressions.  For the interior region we denote these by $\varphi^L_\pm$ and use $A_L$, and $B_L$.   

For $x >0$, the mode function $\varphi_H^R e^{-i \w t}$ has the behavior on the past horizon \cite{fba2016}
%~\cite{fba2016} 
%(note the different normalization with respect to the 2D modes of the previous section, following from (\ref{2d}))
\be \varphi_H^R e^{-i \w t} \to \sqrt{\frac{mv_0}{4 \pi \w n_{1D}\hbar}} e^{-i \w (t - \frac{z}{v_0})}  \;, \ee
and the mode function $\varphi_I e^{-i \w t}$ has the behavior  on past null infinity
\be \varphi_I e^{-i \w t_S} \to  \sqrt{\frac{mc_R}{4 \pi \w n_{1D}\hbar}} e^{-i \w (t + \frac{z}{c_R})} \;. \ee
Therefore, the exact solutions were found to be 
\bes \bea \phi_H^R &=& e^{-i \w t}\sqrt{\frac{mv_0}{4 \pi \w n_{1D}\hbar}} \; \frac{\Gamma(c_+ - a_+) \, \Gamma(c_+ - b_+)}{\Gamma(c_+) \Gamma(c_+-a_+-b_+)} \; \varphi^R_+  \;, \label{phiH-soln} \\ 
\phi_I &=& e^{-i \w t}\sqrt{\frac{mc_R}{4 \pi \w n_{1D}\hbar}} \; \frac{\Gamma(c_- - a_-) \, \Gamma(c_- - b_-)}{\Gamma(c_-) \Gamma(c_--a_+ -b_-)} \; \varphi^R_-  \label{phiI-soln}  \;. \eea \label{phiRI-solns} \ees

In the interior, $z$ is a time coordinate.  There is a solution $ \varphi^{{\rm int}} e^{i \w t}$ that on the past horizon is
\be \varphi^{\rm int} e^{i \w t_S} \to \sqrt{\frac{mv_0}{4 \pi \w n_{1D}\hbar}} e^{-i \w (z/v_0 - t)}  \;. \ee
The exact solution is
\be \phi_H^L = e^{i \w t} \sqrt{\frac{mv_0}{4 \pi \w n_{1D}\hbar}}\, \varphi^L_- \label{varphi-int} \;. \ee

In the interior there are also contributions from $\phi_H^R$ and $\phi_I$. Since they get reflected and transmitted respectively into the interior, they are 
\bes \bea \varphi_H^R e^{-i \w t} &=& %R_H \varphi^{\rm int} R_H  e^{-i \w t}  =
e^{-i \w t}R_H \sqrt{\frac{mv_0}{4 \pi \w n_{1D}\hbar}}\, \varphi^L_- \ ,  \label{phi-H-int}\\
\varphi_I e^{-i \w t_S} &=& %T_I \varphi^{\rm int} e^{-i \w t}
e^{-i \w t_S}T_I \sqrt{\frac{mv_0}{4 \pi \w n_{1D}\hbar}}\, \varphi^L_- \;. \label{phi-I-int}  \eea \ees
Here $R_H$ and $T_I$are the reflection and transmission coefficients for the exterior region which are given by \cite{fba2016}
%~\cite{fba2016}
\bes \bea R_H &=& \frac{\Gamma\left(\frac{i \w}{k v_0}\right) \Gamma\left[\frac{-i \w}{2k}\left(\frac{1}{v_0}+ \frac{1}{c_R} \right)\right] \Gamma\left[1-\frac{i\w}{2k}\left(\frac{1}{v_0}+ \frac{1}{c_R} \right)\right]}{\Gamma\left(\frac{-i \w}{k v_0}\right) \Gamma\left[\frac{-i \w}{2k}\left(\frac{1}{c_R} \frac{1}{v_0} \right)\right] \Gamma\left[1-\frac{i\w}{2k}\left(\frac{1}{c_R}- \frac{1}{v_0} \right)\right]} \;, \label{rh} \\
    T_I &=& \sqrt{\frac{c_R}{v_0}} \; \frac{\Gamma\left[\frac{-i \w}{2k}\left(\frac{1}{v_0}+ \frac{1}{c_R} \right)\right] \Gamma\left[1-\frac{i\w}{2k}\left(\frac{1}{v_0}+ \frac{1}{c_R} \right)\right]}{\Gamma \left(1 - \frac{i \w}{k v_0} \right) \Gamma\left(-\frac{i \w}{k c_R}\right)} \;. \label{ti}
\eea \ees
The other scattering coefficients are
\bea \label{ri} R_I &=&  \frac{\Gamma(\frac{i\omega}{kc_R})}{\Gamma(-\frac{i\omega}{kc_R})}\frac{ \Gamma(-\frac{i\omega}{2k}(\frac{1}{|v|}+\frac{1}{c_R}))\Gamma(1-\frac{i\omega}{2k}(\frac{1}{|v|}+\frac{1}{c_R})) }{ \Gamma(-\frac{i\omega}{2k}(\frac{1}{|v|}-\frac{1}{c_R}))\Gamma(1-\frac{i\omega}{2k}(\frac{1}{|v|}-\frac{1}{c_R})) }\ , \label{ri} \\ 
T_H &=& \sqrt{\frac{|v|}{c_R}}\ \frac{\Gamma(-\frac{i\omega}{2k}(\frac{1}{|v|}+\frac{1}{c_R}))\Gamma(1-\frac{i\omega}{2k}(\frac{1}{|v|}+\frac{1}{c_R}))}{\Gamma(1-\frac{i\omega}{kc_R})\Gamma(-\frac{i\omega}{k|v|})} = T_I\ , \label{th} \\
\alpha &=& 
\sqrt{\frac{|v|}{c_L}}\frac{ \Gamma(1-\frac{i\omega}{k|v|})\Gamma (-\frac{i\omega}{kc_L})}{\Gamma(-\frac{i\omega}{2k}(\frac{1}{c_L}+\frac{1}{|v|}))
\Gamma(1-\frac{i\omega}{2k}(\frac{1}{c_L}+\frac{1}{|v|}))}\ , \label{al} \\
\beta &=& 
\sqrt{\frac{|v|}{c_L}}\frac{ \Gamma(1-\frac{i\omega}{k|v|})\Gamma (\frac{i\omega}{kc_L})}{\Gamma(-\frac{i\omega}{2k}(\frac{1}{|v|}-\frac{1}{c_L}))
\Gamma(1-\frac{i\omega}{2k}(\frac{1}{|v|}-\frac{1}{c_L}))}\ . 
\label{be} \eea

\section{USEFUL INTEGRALS}

The second, fourth and fifth integrals of eq. (\ref{hpdc}) are
\bea  \int_0^\infty d\omega \frac{ \omega^3\cos(\omega\Delta u)}{\sinh^3(\frac{\pi\omega}{\kappa})} &=& \frac{\kappa^6}{16 \pi^2\cosh^4(\frac{\kappa\Delta u}{2})}\Big[ \left(\Delta u^2 + \frac{(6+\pi^2)}{\kappa^2} \right) 
\cosh(\kappa \Delta u) \nonumber \\ &&
+\frac{6}{\kappa^2} -2(\Delta u^2 +\frac{\pi^2}{\kappa^2}) -6\frac{\Delta u}{\kappa}\sinh(\kappa\Delta u)\Big]\ , \\
\int_0^\infty d\omega \frac{ \omega^3\cos(\omega\Delta u)}{\sinh^2(\frac{\pi\omega}{\kappa})}&=& -\frac{3\kappa^4}{4\pi^4}Re \psi''(-\frac{i\kappa\Delta u}{2\pi}) +\frac{\kappa^5}{8\pi^5}
\Delta u\ Im\psi'''(-\frac{i\kappa\Delta u}{2\pi})\ , \\
\int_0^\infty d\omega \frac{ \omega^3 \cosh(\frac{\pi\omega}{\kappa})\cos(\omega\Delta u)}{\sinh^3(\frac{\pi\omega}{\kappa})}&=&- \frac{\kappa^6}{16\pi^2\sinh^4(\frac{\kappa\Delta u}{2})}\Big[ (\frac{6}{\kappa^2}+\Delta u^2)\cosh(\kappa \Delta u) \nonumber \\ && -\frac{6}{\kappa^2}+2\Delta u^2 -6\frac{\Delta u}{\kappa}\sinh(\kappa\Delta u) \Big] \label{HPtp}
\ , \eea
where
\be Re \psi'' (iy)=-2 \sum_{n=1}^\infty \frac{n(n^2-3y^2)}{(n^2+y^2)^3}\ , Im\psi'''(iy)=-24y\sum_{n=1}^\infty \frac{n(n^2-y^2)}{(n^2+y^2)^4}\ . \ee
The integrals in eq. (\ref{scll}) are
\bea  \int_0^\infty d\omega \frac{\omega^2 \cos(\omega(u-v'))}{\sinh^2(\frac{\pi\omega}{\kappa})}&=& \frac{\kappa^3}{4\pi} \frac{\left( -2+(u-v')\kappa \coth(\frac{\kappa(u-v')}{2})\right)}{\sinh^2(\frac{\kappa(u-v')}{2})}|_{u-v'\to 0} \to \frac{\kappa^3}{6\pi}\ , \nonumber \\
\int_0^\infty d\omega \frac{\omega^2 \cos(\omega(u-v'))}{\sinh(\frac{\pi\omega}{\kappa})} &=&  - \frac{\kappa^3}{4\pi^3} Re \psi'' (-\frac{i\kappa(u-v')}{2\pi}+\frac{1}{2})|_{u-v'\to 0} \to
-\frac{\kappa^3 \psi''(\frac{1}{2})}{4\pi^3}%= \frac{\kappa^3}{4\pi^3} (16.8288) 
\ , \nonumber \\ \int_0^\infty d\omega \frac{\omega^2  \cosh (\frac{\pi\omega}{\kappa}) \cos(\omega(u-v'))}{\sinh^2(\frac{\pi\omega}{\kappa})} &=& -\frac{\kappa^3}{4\pi} 
\frac{\left( -2+(u-v')\kappa \tanh(\frac{\kappa(u-v')}{2})\right)}{\cosh^2(\frac{\kappa(u-v')}{2})}|_{u-v'\to 0} \to \frac{\kappa^3}{2\pi} \nonumber \ ,
  \eea
 where $\psi''(\frac{1}{2}) \approx -16.8288$. 
 \end{appendix}

  \end{document}